\documentclass[journal=jctcce,manuscript=article,layout=traditional]{achemso}
\usepackage[utf8]{inputenc}

\usepackage{amssymb,amsmath}  
\usepackage{gensymb}
\usepackage{mathrsfs}         
\usepackage{lmodern}          
\usepackage{graphicx}         
\usepackage{multirow}         
\usepackage{color}            
\usepackage{pifont}           
\usepackage{verbatim}         
\usepackage{svg}              

\usepackage{graphicx}
\usepackage[colorlinks=true, allcolors=blue]{hyperref}
\usepackage{wrapfig}
\usepackage{multirow}
\usepackage{multicol}

\usepackage{color}                                          
\usepackage{soul}                                           

\title{Robust, accurate, and efficient: quantum embedding using the Huzinaga level-shift projection operator for complex systems}

\author{Daniel S. Graham}
\author{Xuelan Wen}
\author{Dhabih V. Chulhai}
\affiliation{Department of Chemistry, University of Minnesota. 207 Pleasant St. SE, Minneapolis MN
55455, USA.}
\alsoaffiliation{Current address: Department of Chemistry, University of Indianapolis. 1400 E Hanna Ave.,
Indianapolis, IN 46227, USA.}
\author{Jason D. Goodpaster}
\email{jgoodpas@umn.edu}
\affiliation{Department of Chemistry, University of Minnesota. 207 Pleasant St. SE, Minneapolis MN 
55455, USA.}
\date{\today}

\begin{document}

\maketitle

\begin{abstract}
Wave function (WF) in density functional theory (DFT) embedding methods provide a framework for performing localized, high accuracy WF calculations on a system, while not incurring the full computational cost of the WF calculation on the full system. In order to effectively partition a system into localized WF and DFT subsystems, we utilize the Huzinaga level-shift projection operator within an absolutely localized basis. In this work, we study the ability of the absolutely localized Huzinaga level-shift projection operator method to study complex WF and DFT partitions, including partitions between multiple covalent bonds, a double bond, and transition metal-ligand bonds.  We find that our methodology can accurately describe all of these complex partitions.  Additionally, we study the robustness of this method with respect to the WF method, specifically where the embedded systems were described using a multiconfigurational WF method.  We found that the method is systematically improvable with respect to both the number of atoms in the WF region and the size of the basis set used, with energy errors less than 1 kcal/mol. Additionally, we calculated the adsorption energy of H$_2$ to a model of an iron metal organic framework (Fe-MOF-74) to within 1 kcal/mol compared to CASPT2 calculations performed on the full model while incurring only a small fraction of the full computational cost.  This work demonstrates that the absolutely localized Huzinaga level-shift projection operator method is applicable to very complex systems with difficult electronic structures.
\end{abstract}

\section{Introduction}

One of the fundamental challenges of quantum chemistry is balancing computational cost and accuracy.   
Kohn-Sham Density Functional Theory\cite{Kohn1965Self-ConsistentEffects, Hohenberg1964InhomogeneousGas} (KS-DFT) has been a computational chemistry mainstay as it balances cost and accuracy well for many chemical systems.
However, KS-DFT relies on an approximate exchange-correlation functional which results in several well documented deficiencies including underestimation of chemical reaction barriers, and inaccurate description of degenerate and near-degenerate states, such as in transition metal systems and covalent bond dissociation.\cite{Cohen2008InsightsTheory,Jones2015DensityFuture,Yu2016Perspective:Staircase,Cohen2012ChallengesTheory, Moltved2019TheConsequences} 
These interactions are essential for accurately describing a variety of systems such as gas binding to metal organic frameworks (MOFs) and reaction energy barriers. 
Correlated wave function (WF) methods such as coupled cluster (CC)\cite{Cizek1966OnMethods, Cizek1971CorrelationMethodst} and complete active space (CAS)\cite{Werner1980ACoefficients} have been shown to more accurately reproduce the aforementioned interactions.\cite{Swart2007EnergyMethods,Rubes2012CombinedMOF} 
Additionally, most WF methods are systematically improvable: the accuracy of calculation may be improved through a well defined process (e.g. including additional excitations for CC calculations, or increasing the size of the CAS active space). 
Yet, for large systems the computational cost of most WF methods are several orders of magnitude larger than KS-DFT methods. 
Frequently, one is interested in chemical transformations that are localized to a small region of the overall system, such as bond formation or elimination, molecular adsorption, or bond rotation.   Many embedding methods such as QM/MM\cite{warshel1976theoretical}, ONIOM\cite{Svensson1996}, DMET\cite{Knizia2013, Sun2016}, embedded mean-field theory\cite{Ding2017EmbeddedPartitioning,Ding2017Linear-ResponseTheory, Miyamoto2016Fock-MatrixTheory, Fornace2015, Fornace2015CorrectionTheory}, Green's function embedding\cite{Onida,Chibani2016, Sun2016}, partition DFT,\cite{Elliott2010PartitionTheory, Nafziger2014Density-BasedCalculations, Gomez2017Partition-DFTDimer} and DFT embedding\cite{Jacob2014, Sun2016} among many others \cite{Hedegard2016PolarizableGroup,Muhlbach2018QuantumLevel, Hrsak2018PolarizableMethod, Culpitt2016MulticomponentFormulation, Nanda2018TheApproach}, take advantage of this intrinsic localization of chemical transformations to achieve substantially improved accuracy for a nominal additional computational cost. 
By dividing the total system into subsystems, important local interactions can be accurately modeled at significantly reduced computational cost.   This approach is particularly advantageous when performing WF calculations, due to the steep computational scaling of WF methods.  


DFT embedding provides a formally exact framework for subdividing a system where the interactions between subsystems are treated using DFT.\cite{Jacob2014, Wesolowski2013, Wesoowski2006One-ElectronSystems, Neugebauer2010, Yang1991, Huang2011, Goodpaster2010} The DFT subsystem interaction potential can then be used to easily embed a WF calculation within the DFT potential of the full system.\cite{cheng2017potential,Goodpaster2012,manby2012simple,libisch2014embedded,govind1998accurate,Govind1999,Hofener2041,Gomes2008,Khait2010,Kluner2001,Kluner2002a,SeveroPereiraGomes2012,Huang2008, Huo2016BreakingCatalyst}
Calculation of the subsystem interaction potential however,  is the central challenge of DFT embedding methods.
The interaction potential for DFT embedding differs from KS-DFT as subdividing a system introduces a non-additive kinetic energy component.
This non-additive kinetic energy may be approximated\cite{Thomas1927TheFields,Fermi1928AElements,Weizsacker1935ZurKernmassen,Fux2008,Beyhan2010,Jacob2014, Ramos2015PerformanceReactions, Pavanello2011ModellingFormalism, Pavanello2013OnDFT, Gotz2009b}, numerically calculated\cite{Jacob2014, Goodpaster2010, Roncero2008AnPotentials}, or eliminated all together through subsystem orbital orthogonalization\cite{Gordon2012,Chulhai2017, Chulhai2015FrozenSystems, Claudino2019AutomaticTheories, Khait2012OnTheory, Tamukong2014DensityOrthogonality, Culpitt2017Communication:Procedure}.
The use of subsystem orbital orthogonalization methods for exact DFT embedding was studied by the Manby and Miller groups through the use of a constant shift $\mu$-projection operator\cite{manby2012simple, Chapovetsky2018PendantReduction, Goodpaster2012, Goodpaster2014AccurateWavefunctions, Welborn2018Even-handedEmbedding}.
This $\mu$-projection operator demonstrated impressive results and in a later paper, Kallay and co-workers suggested\cite{Hegely2016ExactLocalization} the use of the  Huzinaga\cite{Huzinaga1971TheorySystems,Francisco1992GeneralizedTheory, Olsen2015PolarizableStrategy} level-shift projection operator as an alternative to the $\mu$-projection operator. 
Our group generalized the Huzinaga level-shift projection operator with a freeze-and-thaw localization scheme and demonstrated significant success using absolute localization on molecular and periodic systems.\cite{Chulhai2017, Chulhai2018Projection-BasedSystems} 

In order to make WF-in-DFT embedding feasible for large systems, the number of valence orbitals in the WF region must be managed. 
Including the basis functions of the full system in the embedded WF subsystem simply moves orbitals not occupied in the WF subsystem to the virtual space, which for CC calculations actually increases the computational cost upon embedding due to the higher scaling of CC methods with respect to virtual orbitals compared to occupied orbitals.
There have been several methods proposed for reducing the number of basis functions in the WF subsystem while maintaining high accuracy such as basis set truncation\cite{Bennie2015, Barnes2013AccurateEmbedding}, ``bottom up" basis set extension\cite{Bensberg2019AutomaticEmbedding}, dual basis set approximation\cite{Hegely2018DualSchemes}, and concentric localization and truncation of virtual space\cite{Claudino2019SimpleLocalization}.
Our absolute localization scheme may be considered the most strict of the truncation methods mentioned, including only basis functions in the WF calculation which are centered on the atoms specified for the WF subsystem. While this constraint does increase the errors for absolute energies, we have found that absolute basis truncation actually performs better for WF-in-DFT reaction energies than using the full system basis. \cite{Chulhai2017} We argue this improved reaction energy is due to a systematic cancellation of error between products and reactants enabled by the strict localization of the WF orbitals. 

Expanding upon our previous study, the following results indicate that this method provides robust, systematically improvable results for a diverse test set, and thus has applicability to a wide variety of chemical systems.
These systems include partitioning across two covalent bonds and a conjugated $\pi$ network, and gas adsorption onto transition metals among others.
Additionally, we provide recommendations for system partitioning to achieve the highest degree of chemical accuracy. Finally, we demonstrate the power of this method for multiconfigurational embedding by applying it to a MOF cluster system approaching the limit of traditional WF methods.

\section{Theory}
In subsystem DFT methods, the electron density, $\gamma$, of a system is subdivided into two subsystems, 
\begin{equation}
    \gamma = \gamma^{\text{A}} + \gamma^{\text{B}}
\end{equation}
where $\gamma^{\text{A}}$ and $\gamma^{\text{B}}$ are the electron densities of subsystem $\text{A}$ and $\text{B}$ respectively.
Level shift projection operators enforce orthogonality via adding a projection operator to the subsystem Hamiltonian. 
 Because of the general form, the embedding and projection potentials may be added to any subsystem Hamiltonian to calculate energy, either for DFT or WF theory methods. 
The Huzinaga projection operator is,
\begin{equation}
   \mathbf{P}^\text{B} = - \frac{1}{2} \left( \mathbf{F}^\text{AB} \gamma^\text{B} \textbf{S}^\text{BA} + \textbf{S}^\text{AB} \gamma^\text{B} \mathbf{F}^\text{BA} \right)
   \label{eq:huzinaga-operator}
\end{equation}
where $\mathbf{F}^{AB(BA)}$ are the AB (or BA) block of the total Fock matrix and $\mathbf{S}^\text{AB(BA)}$ are the AO overlap matrices between subsystems A and B. Thus, for subsystem A, orbitals within A that are not orthogonal to subsystem B are shifted to higher energies, and \emph{vice versa}. 

The projected Fock matrix of subsystem A embedded in subsystem B is,
\begin{equation}
   \mathbf{f}^\text{A-in-B} = \mathbf{h}^\text{A-in-B}[\gamma^{\text{A}}, \gamma^{\text{B}}] + \mathbf{J}[\gamma^\text{A}] + \mathbf{v}_\text{xc}[\gamma^\text{A}]
\end{equation}
where $\mathbf{J}$ is the electron Coulomb potential, $\mathbf{v}_\text{xc}$ is the exchange-correlation (XC) potential, and the embedded core Hamiltonian is
\begin{equation}
    \mathbf{h}^\text{A-in-B}[\gamma^\text{A}, \gamma^\text{B}] = \mathbf{h} + \mathbf{J}[\gamma^\text{A}+ \gamma^\text{B}]
    - \mathbf{J}[\gamma^\text{A}]
    + \mathbf{v}_{\text{xc}}[\gamma^\text{A}+ \gamma^\text{B}]
    - \mathbf{v}_{\text{xc}}[\gamma^\text{A}] 
    +  \mathbf{P}^\text{B},
       \label{eq:embeddedcore}
\end{equation}
where $\mathbf{h}$ is the total one-electron Hamiltonian.

The form of our overall WF-in-DFT embedding energy,
\begin{equation}
    E_{\text{WF-in-DFT}}^{\text{Full}} = E_{\text{KS-DFT}}^{\text{Full}} - E_{\text{DFT}}^{\text{A}} + E_{\text{WF}}^{\text{A}},
\end{equation}
uses an subtractive embedding framework like ONIOM\cite{Svensson1996}, where $E_{\text{WF-in-DFT}}^{\text{Full}}$ is the total WF-in-DFT energy, $E_{\text{KS-DFT}}^{\text{Full}}$ is the canonical KS-DFT of the full system, $E_{\text{DFT}}^{\text{A}}$ is the DFT energy of subsystem $\text{A}$ embedded in the full system, 
\begin{equation}
    E_{\text{DFT}}^{\text{A}} = \text{Tr}\left( \gamma^\text{A} \cdot \mathbf{h}^\text{A-in-B}[\gamma^\text{A}, \gamma^\text{B}] \right)  +  \text{J}[\gamma^\text{A}] + E_\text{xc}[\gamma^\text{A}],
\end{equation}
and $E_{\text{WF}}^{\text{A}}$ is the WF energy of subsystem $\text{A}$ embedded in the DFT potential of the full system
\begin{equation}
    E_{\text{WF}}^{\text{A}} = \langle \Psi^\text{A} | \hat{H}^\text{A-in-B} | \Psi^\text{A} \rangle.
\end{equation}
This form differs from several previous energy formulations\cite{Goodpaster2012, Bennie2015} by correcting using the fully relaxed, total KS-DFT energy of the system in a similar manner to Carter et. al.\cite{Huang2011} 

\section{Computational Details}

All organic molecule geometries were optimized using Gaussian 16\cite{g16} with the M06 functional\cite{Zhao2008} and aug-cc-pVTZ basis\cite{Dunning1989GaussianHydrogen}, and are reported in the SI. The Fe-MOF-74 cluster geometry was calculated using Gaussian 16\cite{g16} following the procedure outlined by Lee and coworkers\cite{Lee2014Design4} and included in the SI.
Single point DFT and CCSD(T) energy calculations were calculated using PySCF version 1.6\cite{Sun2018PySCF:Framework} and CASPT2 calculations were done using Molpro 2019.2\cite{Werner2012Molpro:Package, MOLPRO2019, Gyorffy2013AnalyticalFitting, Shiozaki2011Communication:Gradients, Celani2000MultireferenceFunctions}.
Embedded CASPT2 results were calculated using Molpro 2012.1\cite{Werner2012Molpro:Package,Gyorffy2013AnalyticalFitting, Shiozaki2011Communication:Gradients, Celani2000MultireferenceFunctions}.
For all DFT calculations, PySCF grid level was set to 4.

In order to generate a final WF-in-DFT embedding energy, our method utilizes a freeze and thaw\cite{Wesolowski1996b} scheme and an ONIOM\cite{Svensson1996} energy formulation; this method is as follows:
(1) We start with a full system KS-DFT calculation.
(2) The resulting converged full system KS-DFT density is used as the initial density guess for subsystem density.
To generate the subsystem density guess in the absolutely localized basis, the components of the full system density matrix consisting of basis functions centered on subsystem atoms are extracted and normalized to create an initial subsystem density matrix. 
(3) The subsystem density is then relaxed following a freeze and thaw protocol, where one subsystem density is allowed to relax while the rest of the system density is frozen, then that subsystem is frozen and another subsystem is ``thawed'' and allowed to relax.
During freeze and thaw relaxation, the full system Fock matrix, used in the projection operator, is recomputed each time all subsystems undergo a relaxation cycle.
(4) Finally, once the subsystem densities are determined converged between freeze and thaw cycles, embedded WF calculations may be performed using DFT-in-DFT potentials. 
WF calculations use the DFT subsystem density as a starting guess for an initial Hartree-Fock calculation followed by the subsequent WF calculation and utilize the embedded core Hamiltonian as defined in eq. \ref{eq:embeddedcore}. All embedding calculations were performed with our open source Quantum Solid state and Molecular Embedding (QSoME) code \cite{qsome2019}.

\section{Results and Discussion}

\subsection{Systematic Improvability}

A useful feature of many WF methods is their systematic improvability --- there is a clearly defined procedure for improving results towards the exact solution. 
One direction of systematic improvability is the description of the wave function; for instance coupled cluster methods are improved by including additional excitations and multiconfigurational methods are improved by increasing the number of configurations.  As our method is generality applicable to any WF method, we know that we can always systematically improve the quality of the wave function.  Furthermore, we have previously used \ensuremath{i-}FCIQMC\ in our embedding method;\cite{Petras2019FullyCarlo} therefore, our embedding scheme is compatible with other high-quality WF methodologies.
In addition to the systematic improvability of the description of the wave function, the basis set can also be expanded to converge to the complete basis set limit.  In practice, one typically does calculations at increasing basis set size and extrapolates to the complete basis set limit.  It is not obvious that absolutely localized  projection-based embedding will allow for the same extrapolation schemes.  Additionally, compared to WF calculations performed on the full system, embedding calculations have an additional convergence criteria, which is the size of the WF subsystem.    In order to determine the systematic improvability of the Huzinaga level-shift embedding method, embedding results were compared to full system WF energies while varying the size of the WF subsystem and the number of basis functions. 
We show that Huzinaga WF-in-DFT embedding is systematically improvable for increasing number of atoms in the WF subsystem, and increasing number of basis functions describing the system.

\begin{figure}
    \centering
    \includegraphics[width=.5\textwidth]{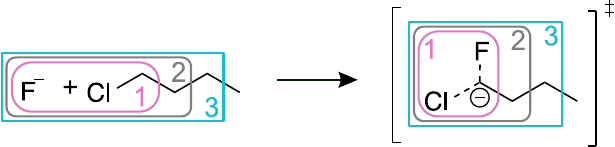}
    \caption{1-chlorobutane $\text{S}_{\text{N}}\text{2}$ transition state reaction. Numbering indicates the size of WF subsystem, where the number corresponds to the number of carbon atom centers in the subsystem. The smallest WF subsystem includes only the carbon center most local to the $\text{S}_{\text{N}}\text{2}$ reaction. For other $\text{S}_{\text{N}}\text{2}$ reactions studied, WF subsystems similarly incrementally increase.}
    \label{fig:sn2_rxn}
\end{figure}

\begin{figure}
 \includegraphics[width=0.5\textwidth]{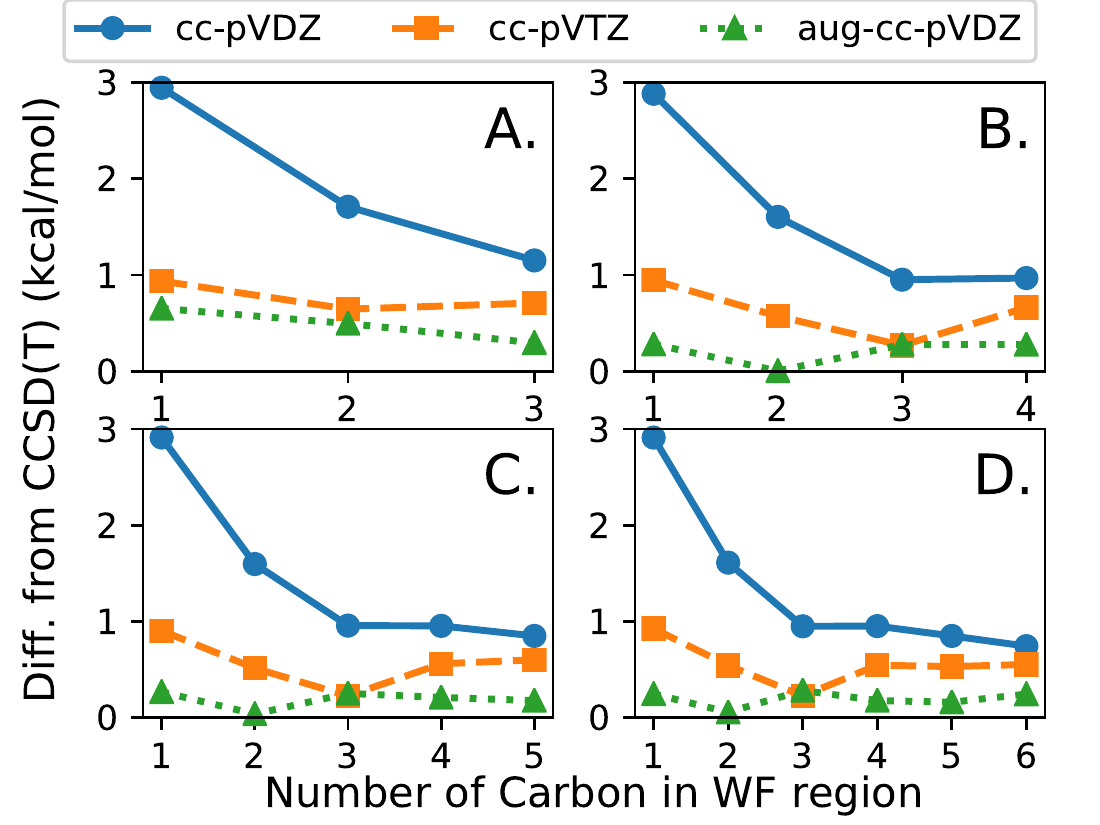}
 \caption{Absolute energy difference of WF-in-DFT embedding from full system CCSD(T) in the same basis set of $\text{S}_{\text{N}}\text{2}$ activation energy. Reactants are, (A) 1-chlorobutane, (B) 1-chloropentane, (C) 1-chlorohexane, and (D) 1-chloroheptane. Number of carbon in the WF subsystem corresponds to the WF subsystem subdivisions represented in figure \ref{fig:sn2_rxn}.}
 \label{fig:sn2_diff}
\end{figure}

The particular reactions we studied to determine systematic improvability were a series of $\text{S}_{\text{N}}\text{2}$ activation energies. 
The $\text{S}_{\text{N}}\text{2}$ activation energy was studied in a previous article from our group\cite{Chulhai2017} and demonstrated the success of the embedding method when dividing a system across a single covalent bond. 
However, where the antecedent article only studied one system subdivision for the reaction, here the system is divided into subsystems of increasingly large WF subsystem sizes. 
These subsystem divisions are shown in Figure \ref{fig:sn2_rxn}, where the size of the WF subsystem increases by including carbon centers along the alkane chain. 
The smallest WF subsystem includes only carbon 1, the halogen atoms, and the hydrogen bonded to carbon 1, as these are the atoms closest to the region of chemical change in the system.
The WF subsystem is increased in size by including the carbon and its bonded hydrogen directly adjacent to the WF subsystem.
Subsystems were charged to create closed shell fragments with WF subsystem given an additional -1 charge and DFT subsystem given an additional +1 charge. 
WF-in-DFT energies were calculated using CCSD(T) as the WF method and M06\cite{Zhao2008} as the DFT exchange-correlation functional. 
Four halogenated hydrocarbon systems energies were calculated using the incrementally increasing WF subsystem method described above.
These embedding energies using the specified  basis set were then compared to the CCSD(T) energy of the corresponding full system in the same basis set (Figure \ref{fig:sn2_diff}).  

Using the cc-pVDZ basis set, WF-in-DFT energies systematically converged to the full system CCSD(T) energy with increasing size of the WF region to within 1 kcal/mol. 
For all alkane systems, energies consistently approach full system results until convergence of the energy with the inclusion of 3 carbon in the WF subsystem. 
Not only do energies converge with increasing WF subsystem size, but WF-in-DFT energies converge irrespective of total size of the system.
Since convergence is not dependent on the total size of the system, the size of the WF calculation may be limited to a much smaller subset of the entire system.
We then tested how increasing the number of basis functions affected convergence.
From the data in Figure \ref{fig:sn2_diff} basis sets larger than cc-pVDZ converge with even fewer carbon atoms in the WF region. 
Therefore, these results demonstrate that by increasing the number of atoms in the WF region the calculations converge to the result obtained from the full WF calculation, regardless of the size of the basis.

\begin{figure}
\includegraphics[width=0.5\textwidth]{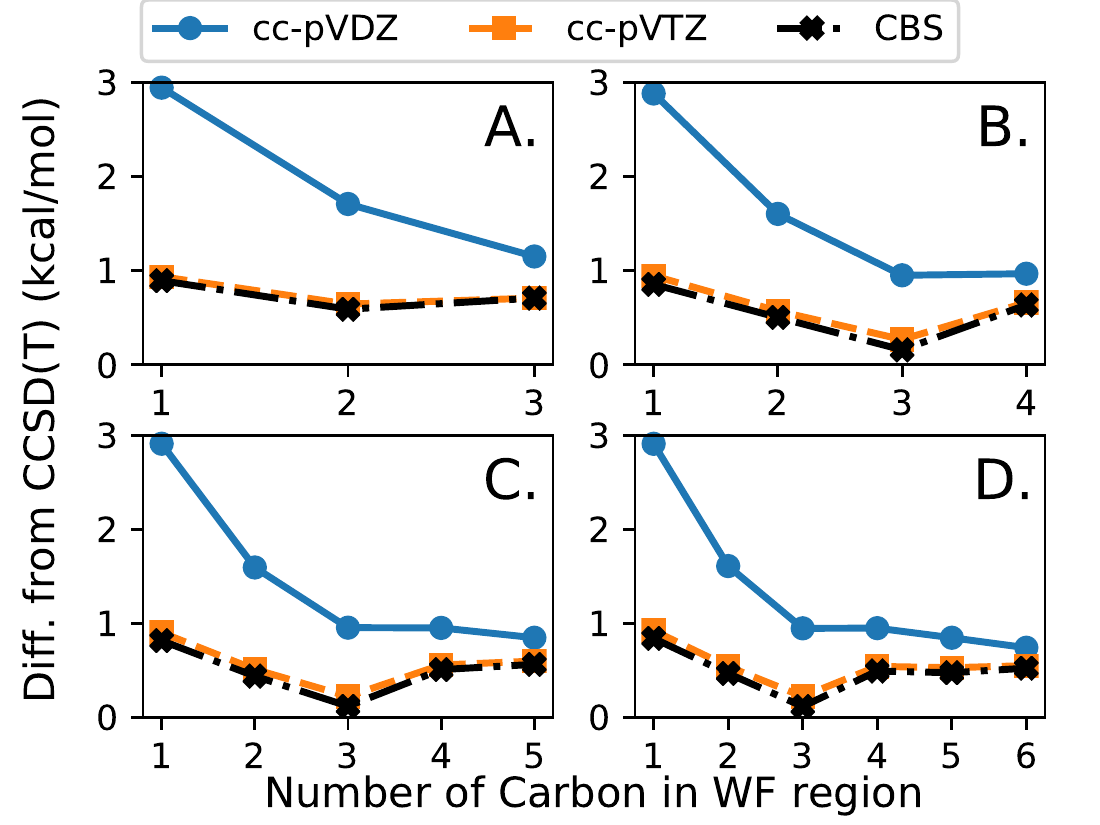}
\caption{Absolute energy difference of $\text{S}_{\text{N}}\text{2}$ activation with complete basis extrapolation\cite{Helgaker1997Basis-setWater,Halkier1998Basis-setH2O}. Reactants are (A) 1-chlorobutane, (B) 1-chloropentane, (C) 1-chlorohexane, and (D) 1-chloroheptane.}
\label{fig:cbs_sn2}
\end{figure}

Next, we tested the accuracy of embedding for an extrapolation to the complete basis set (CBS) limit using cc-pVDZ and cc-pVTZ energies within the formula of Helgaker et al.\cite{Helgaker1997Basis-setWater, Halkier1998Basis-setH2O}.
To do this, we replaced the WF correlation energy of the cc-pVTZ WF-in-DFT calculations with that from the CBS extrapolation of the WF region and compared these extrapolated WF-in-DFT results to the CSB extrapolated WF energy of the full systems.
The extrapolated results, shown in Figure \ref{fig:cbs_sn2}, lie very close to the WF-in-DFT cc-pVTZ results, which indicates that the extrapolation does not change the difference between WF-in-DFT and the full system calculation significantly.   
The similarity between cc-pVTZ and CBS energies for this system supports the use of basis set extrapolation methods within our embedding framework.

WF-in-DFT energies are shown to converge to within 1 kcal/mol of the full system WF energy with both increasing size of WF subsystem, and increasing number of basis functions used to describe the system. 
In both cases, these parameters may be thought of as increasing the space of the WF region, in terms of number of atoms, or degrees of electron freedom, respectively. 
These results demonstrate the systematic improvability of the Huzinaga embedding method.

\subsection{Complex Subsystem Divisions}\label{sec:complex}

We have shown that the Huzinaga level-shift embedding method is systematically improvable and can closely recreate full system WF energies when subsystems are divided across a single covalent bond.
However, in order to be broadly applicable to a variety of systems, the embedding method must be able to handle a myriad of interactions between subsystems.
Embedding methods are often limited by how well the method can treat the interaction between subsystems.
Here we demonstrate the robustness of the Huzinaga embedding method by dividing a system across two distinct covalent bonds and subdividing across a delocalized double bond.

\begin{figure}
    \centering
    
    \includegraphics[width=0.5\textwidth]{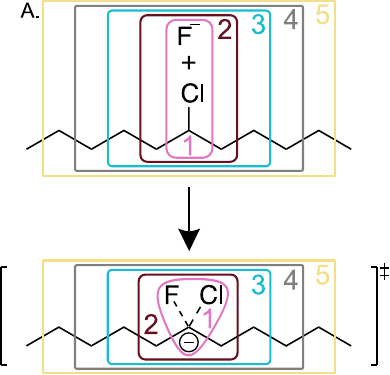}
    \vspace{8px}
    \includegraphics[width=0.5\textwidth]{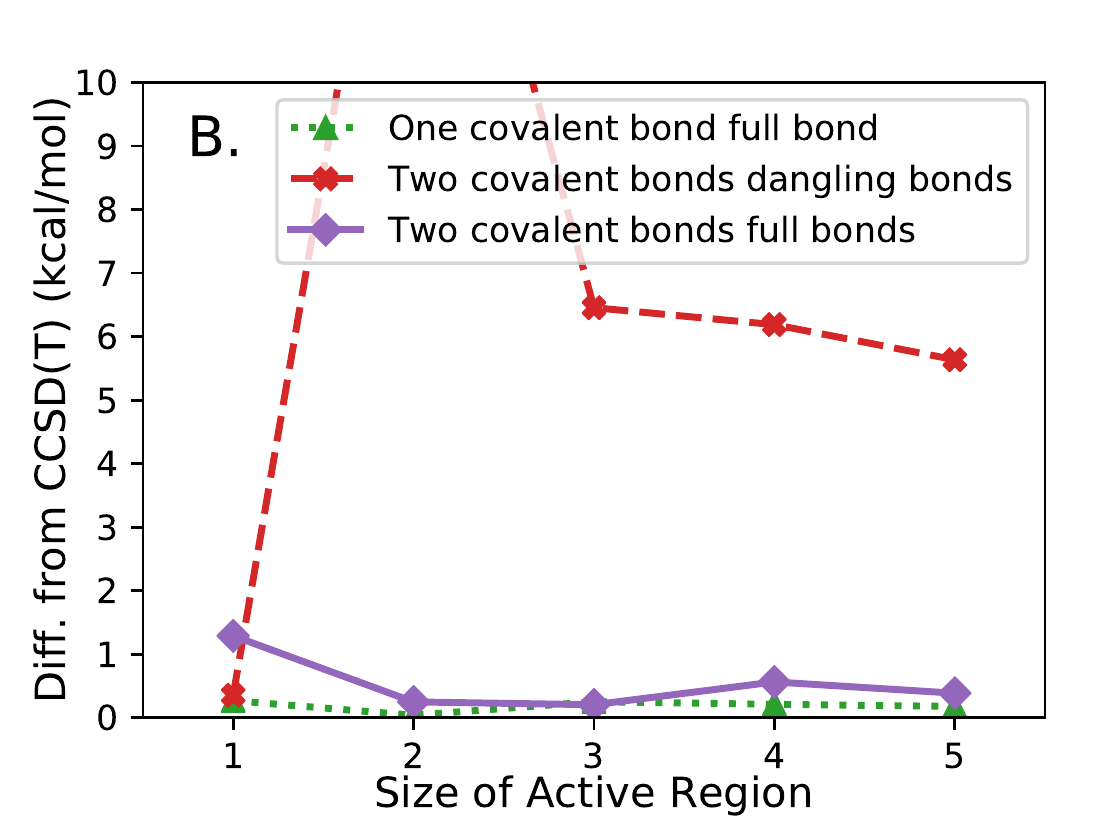}
    \caption{Subfigure (A) 6-chloroundecane reacting to form $\text{S}_{\text{N}}\text{2}$ transition state. Numbering indicates the size of the WF subsystem, where numbers identify the number of carbon away from the reaction in one direction along the chain. This numbering is to provide an analogous measurement to primary carbon $\text{S}_{\text{N}}\text{2}$ shown in Figure \ref{fig:sn2_rxn}.
    Subfigure (B)
    $\text{S}_{\text{N}}2$ activation energy of primary and secondary carbons. Secondary carbon $\text{S}_{\text{N}}2$ reaction energies are shown where the subsystems are charged to include the electrons in the  bond between subsystems within the WF region, and to not include those electrons in the WF subsystem. WF subsystem sizes for partitioning across a single bond are shown in Figure \ref{fig:sn2_rxn}.}
    \label{fig:multicut}
\end{figure}

When subdividing a system, the resulting subsystems are defined by the atoms and electrons included in the subsystem. 
As such, when dividing a complex system into closed shell subsystems, there are a variety of possibilities for the electron distribution among subsystems.
Practically, electrons are distributed among subsystems by specifying the number of electrons associated with each subsystem such that the total number of electrons add up to the total number of electrons in the full system.
When subsystems are not covalently bound, it is reasonable to partition the subsystems such that they are charge neutral, for instance, two water molecules would have 10 electrons each.  
However, when partitioning across a covalent bond, the choice is less obvious.  
One possibility is to partition such that each fragment is closest to charge neutral. 
For instance, in  Figure \ref{fig:multicut}A in partition 1, one subsystem has a fluorine anion (10 electrons), chloride atom (17 electrons), carbon (6 electrons), and hydrogen (1 electron) for a total of 34 electrons and a charge of -1.  
Therefore, one option would be to include 34 electrons; however, this would lead to ``dangling bonds" as the electrons in the bonds between subsystems are split between the WF and DFT subsystems. 
Thus, the other option would be to include 36 electrons and a charge of -3 such that the bonding electrons are included into the WF subsystem.  We refer to this as ``full bonds" partitioning. 
We reiterate for the results presented Figure \ref{fig:cbs_sn2}, to create closed shell subsystems we were forced to subdivide using ``full bonds."

We tested these two options by expanding upon our previous $\text{S}_{\text{N}}2$ results, by studying the activation energy of an $\text{S}_{\text{N}}2$ reaction centered on a secondary carbon.
When embedding this reaction in a WF subsystem, the system was subdivided following the scheme outlined in Figure \ref{fig:multicut}A.
The chosen subdivisions allow for direct comparisons to the previous primary $\text{S}_{\text{N}}2$ reaction results, while also demonstrating how partitioning across multiple bonds affect overall embedding energies.
The WF method and DFT exchange-correlation functional were the same as the primary carbon $\text{S}_{\text{N}}2$ reaction embedding: CCSD(T) and M06\cite{Zhao2008}, respectively. 
The basis used to generate the data in Figure \ref{fig:multicut}B was aug-cc-pVDZ.

Figure \ref{fig:multicut} shows the results of various subsystem charging strategies. 
The ``One covalent bond full bond'' results (green dotted line) are those of the hexane $\text{S}_{\text{N}}2$ system in the previous section (Figure \ref{fig:cbs_sn2}) with an aug-cc-pVDZ basis.
The partitioning across two single covalent bonds results both follow the subdivision scheme outlined in Figure \ref{fig:multicut}A, but the ``Two covalent bonds dangling bonds'' results (red dashes) use a neutral charge DFT subsystem and a -1 charge WF subsystem, while the ``Two covalent bonds full bonds'' results (purple line) show embedding when the WF subsystem has a -3 charge (1 electron added from each dangling bond) and the DFT subsystem has a +2 charge. 
From the data in Figure \ref{fig:multicut}B, it is clear that some subsystem charging is necessary to maintain systematic improvability and significant similarity to full system WF results.
We note, this strategy is similar to adding capping hydrogen atoms in a QM/MM calculation to prevent dangling bonds\cite{Singh1986APolyethers}.
Once properly charged, subdividing a system across multiple covalent bonds gives similar results to subdividing a system across a single covalent bond, indicating that the accuracy of embedding energy is independent of number of bonds which connect the subsystems.
After reviewing the electron density plots of the different subsystem charging strategies (Figure S1 and Figure S2), we can see that including the electrons in the bond between subsystems in the WF subsystem localizes that bond within the WF subsystem. 

The importance of including the electrons in the bond between the subsystems within the WF region may also be seen with the accuracy of the augmented basis sets.
For most of the systems studied, augmented basis sets require the fewest atoms in the WF region to achieve energies below 1 kcal/mol of the full system WF calculation.
We attribute this to the augmented basis functions accommodating the electrons in the bond between the subsystems and having additional flexibility to recreate these bond orbitals.
From these results it would seem as though the more freedom the electron density has to recreate the KS-DFT electron density, the better the reaction energies. 
A natural conclusion would be to use the supersystem basis, rather than the absolutely localized basis for improved accuracy in WF-in-DFT reaction energies. 
In a previous paper, however, our group demonstrated higher accuracy for absolutely localized WF-in-DFT reaction energies than for supersystem basis WF-in-DFT reaction energies.\cite{Chulhai2017} 
We attribute this improved accuracy to systematic error cancellation.
When comparing the DFT-in-DFT energy of subdividing a system across multiple covalent bonds system to the fully relaxed KS-DFT energy, the role of error cancellation seems to be an important factor (Tables S1 and S2).
For the undecane system with subsystem sizes 4 and 5, the absolute DFT-in-DFT energy of the dangling bonds partitioning is closer to the absolute KS-DFT than the full bond partitioning of the same sizes.
However, for those system sizes the full bond partitioning DFT-in-DFT reaction energies are closer to the KS-DFT reaction energies than the dangling bond partitioning.
While the DFT-in-DFT electron density in the aforementioned dangling bond partitioning is closer to the KS-DFT electron density, the systematic error in the electron density of the full bond partitioning enables cancellation of errors and better overall reaction energies.

This suggests that \emph{error cancellation}, and not the ability to most accurately reproduce KS-DFT, dominates the accuracy of WF-in-DFT energy differences.  In the extreme limit where the electrons treated at the WF level of theory are entirely different between products and reactants (for instance, core electrons in the product and valence electrons in the reactants), one would not expect accurate WF-in-DFT energy differences.  Therefore, the opposite limit, where the electrons treated at the WF level of theory are the most similar between products and reactants is likely to produce the most accurate WF-in-DFT energy differences.  This is precisely what absolute localization forces, the electrons are forced to localize on the atoms associated with the subsystem.  Our results for the full bond partitioning are thus suggestive that the electrons in the bonding orbital between subsystems localize similarly between products and reactants, and this leads to better error cancellation and additional accuracy in WF-in-DFT energy differences.  Thus we hypothesize there is an important balance between including enough of the system to encompass the region of interest, demonstrated by the convergence with size of subsystem, and maintaining good error cancellation, by consistently localizing the electron density in the WF region. 
Work in our group is ongoing to further elucidate the role of error cancellation for accurate WF-in-DFT reaction energies.

\begin{figure}
    \centering
    \includegraphics[width=.5\textwidth]{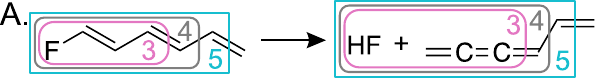}
    
    \vspace{8px}
    \includegraphics[width=0.5\textwidth]{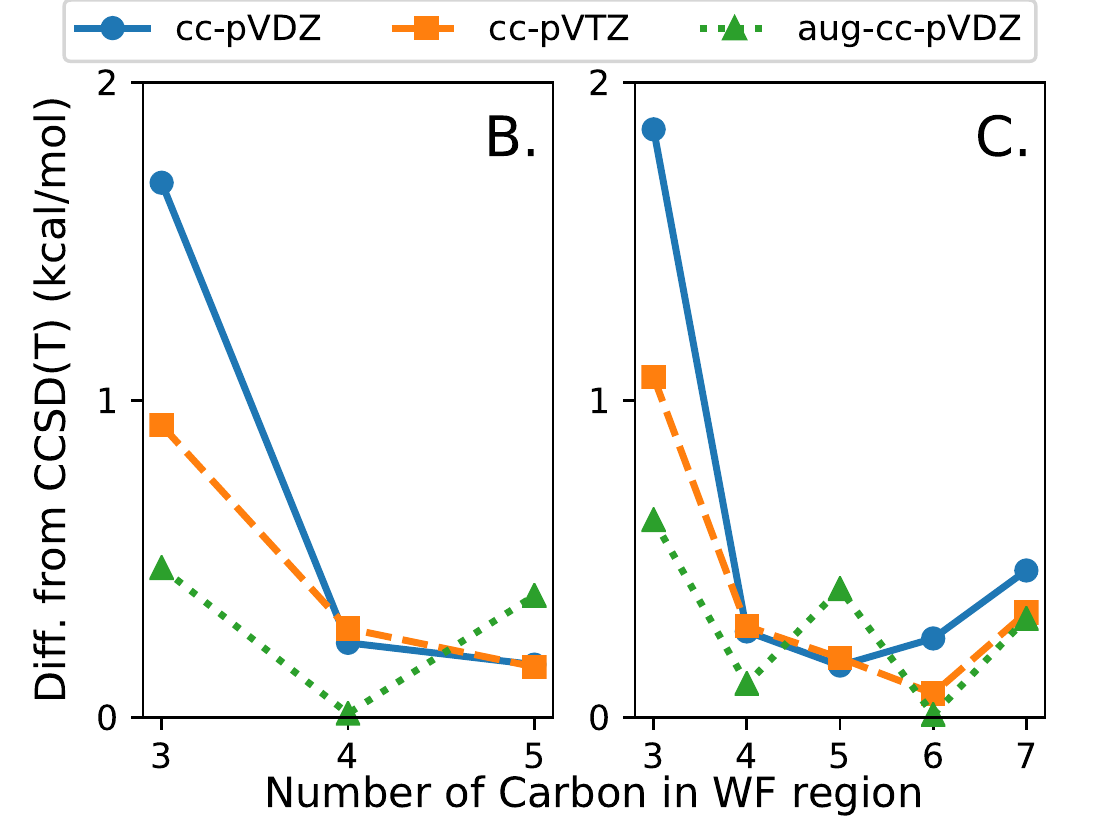}
    \caption{ (A) Flourine elimination reaction. WF subsystem
     sizes are specified following a similar scheme to Figure \ref{fig:sn2_rxn}. Here WF subsystem starts at 3, because 3 carbon undergo bonding changes during the reaction. Subfigures (B) and (C) show absolute energy difference of WF-in-DFT embedding from full system CCSD(T) for flourine elimination reaction. Reactants are (B) (1Z,3E)-1-fluorohexa-1,3,5-triene; and (C) (1Z,3E,5E)-1-fluoroocta-1,3,5,7-tetraene.}
    \label{fig:elim}
\end{figure}

To further demonstrate the robustness of our embedding method, Huzinaga WF-in-DFT energies were calculated for a fluorine elimination reaction. 
This molecule, shown in Figure \ref{fig:elim}A, has a delocalized conjugated $\pi$ system spanning the molecule.
Reaction energies were calculated using CCSD(T) for WF subsystem, and M06\cite{Zhao2008} exchange-correlation functional for the DFT subsystem.  
The subsystems were charged to include the electrons in the bonds between subsystems entirely within the WF region which entailed a -2 charge when partitioning across a double bond and -1 charge when partitioning across a single bond.
The results of embedding across a delocalized system (Figure \ref{fig:elim}), are largely similar to those of the single covalent bond partitioning: the embedding method is systematically improvable with respect to the size of WF subsystem and number of basis functions, in addition to converging below 1 kcal/mol of the full system WF energy.
Therefore, the Huzinaga WF-in-DFT Embedding scheme can handle delocalized $\pi$-bonding networks with the same accuracy as localized covalently bonded systems. 

Complicated systems --- for example, systems that need to divide across many bonds or across delocalized orbitals --- have the potential to introduce additional errors in embedding methods.
However, we have shown that the Huzinaga level-shift projection operator still performs as well for these more complicated systems as it does for simpler single covalent bond partitioned systems, further demonstrating the robustness of the method.
We have shown that as long as the subsystems are divided such that the electrons in the bonds between subsystems are all included in the WF subsystem, then the results are accurate independent of the number of bonds or if the bond is a single or double bond.
In all of these cases, the most important factor for obtaining accurate embedded WF energies is the size of the WF subsystem.

\subsection{Multireference Embedding}

Some of the most challenging systems for quantum chemistry are those with multireference character. 
Systems with degenerate orbitals or partially occupied states often require a multiconfigurational WF method to accurately describe. 
Multiconfigurational WF methods typically scale poorly with size of the system, oftentimes scaling exponentially with the number of electrons and basis functions. However, through absolutely localized Huzinaga WF-in-DFT embedding the WF subsystem includes only a fraction of the total system and therefore a fraction of the total electrons and basis functions. 
By localizing the multireference calculation to a subsystem, we demonstrate multireference energies for systems near the size limit of current non-embedded multireference methods with a greatly reduced computational cost.
$\mu$-projection based embedding has been successfully applied to large multireference systems\cite{Chapovetsky2018PendantReduction,deLimaBatista2017PhotophysicalApproach}, here we demonstrate the accuracy and applicability of the Huzinaga projection operator.

\subsubsection{Homolytic Bond Cleavage}

\begin{figure}
    \centering
    \includegraphics[width=0.5\textwidth]{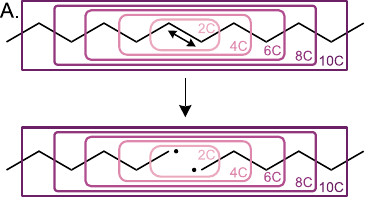}
    
    \vspace{10px}
    \includegraphics[width=0.5\textwidth]{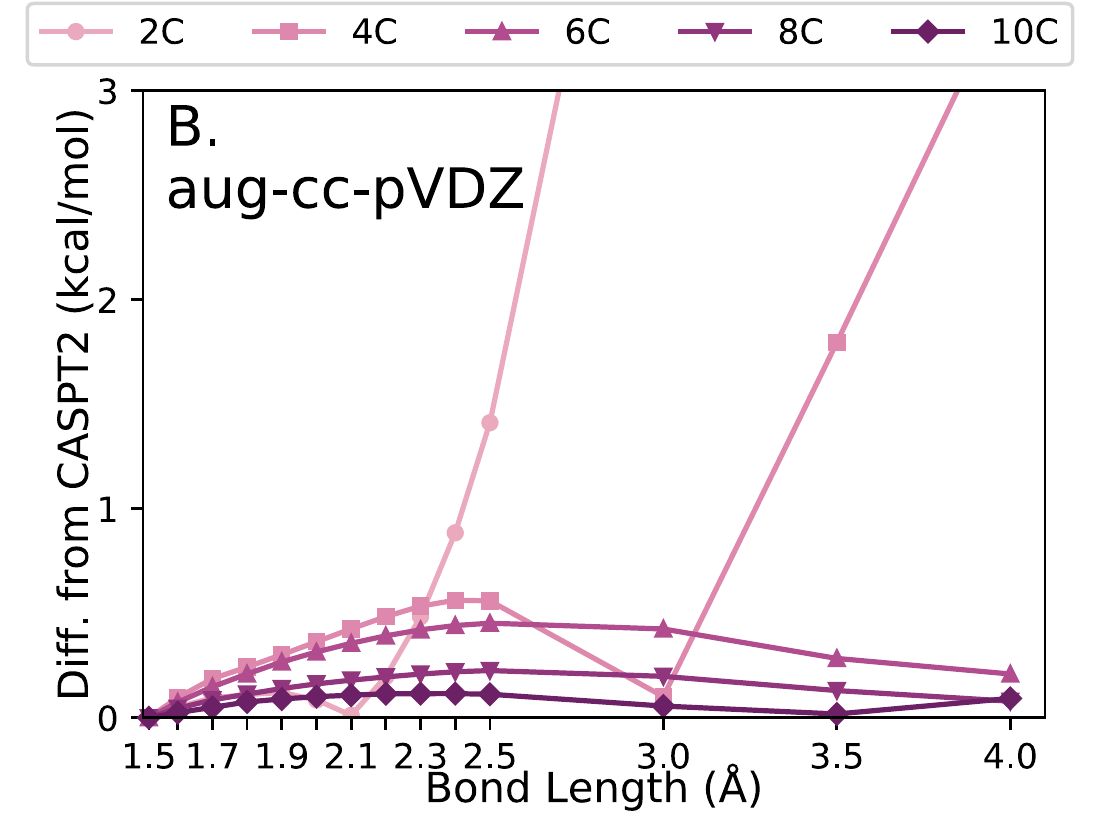}
    
    \vspace{8px}
    \includegraphics[width=0.5\textwidth]{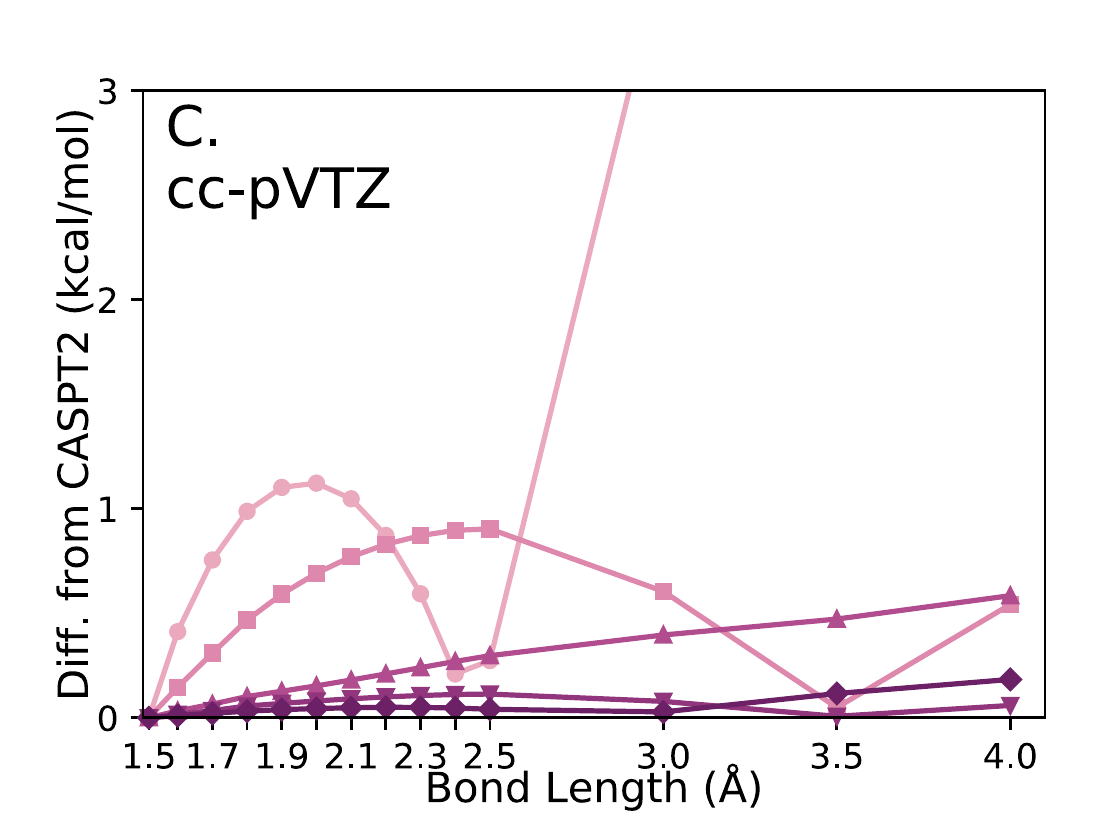}
    \caption{Absolute energy difference of homolytic bond dissociation calculated using CASPT2 embedded in M06\cite{Zhao2008} DFT exchange-correlation functional. Sub-figures show: (A) WF subsystem division; (B) results of embedding with aug-cc-pVDZ; and (C) cc-pVTZ basis.}
    \label{fig:bond_cleavage}
\end{figure}

A relatively simple multireference reaction involves the homolytic bond cleavage of a carbon-carbon bond. 
As the bond elongates, the doubly occupied bonding orbital becomes two degenerate radical orbitals. 
Thus this system provides a good benchmark for determining how well the absolutely localized Huzinaga embedding method can embed a multireference WF method and similar systems have been used previously to benchmark embedded multireference methodologies.\cite{Coughtrie2018EmbeddedTheory}
Using the subsystem charging scheme from the previous section, we embedded a  CASPT2 subsystem within a DFT potential utilizing the M06\cite{Zhao2008} functional (Figure \ref{fig:bond_cleavage}). 
The results demonstrate similar desirable convergence behavior as all previous calculations: increasing the number of carbon in the WF region improves the energies until convergence. 
Additionally, relatively few carbon are necessary in the WF region in order to calculate full system WF level results. 
For those systems with fewer than 6 carbon in the WF region, results significantly differ from multireference energies only in regions of high multireference character.  

\subsubsection{Double Bond Rotation}

\begin{figure}
    \centering
    \includegraphics[width=0.5\textwidth]{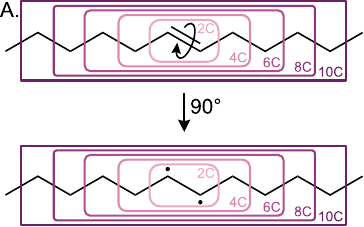}
    
    \vspace{10px}
    \includegraphics[width=0.5\textwidth]{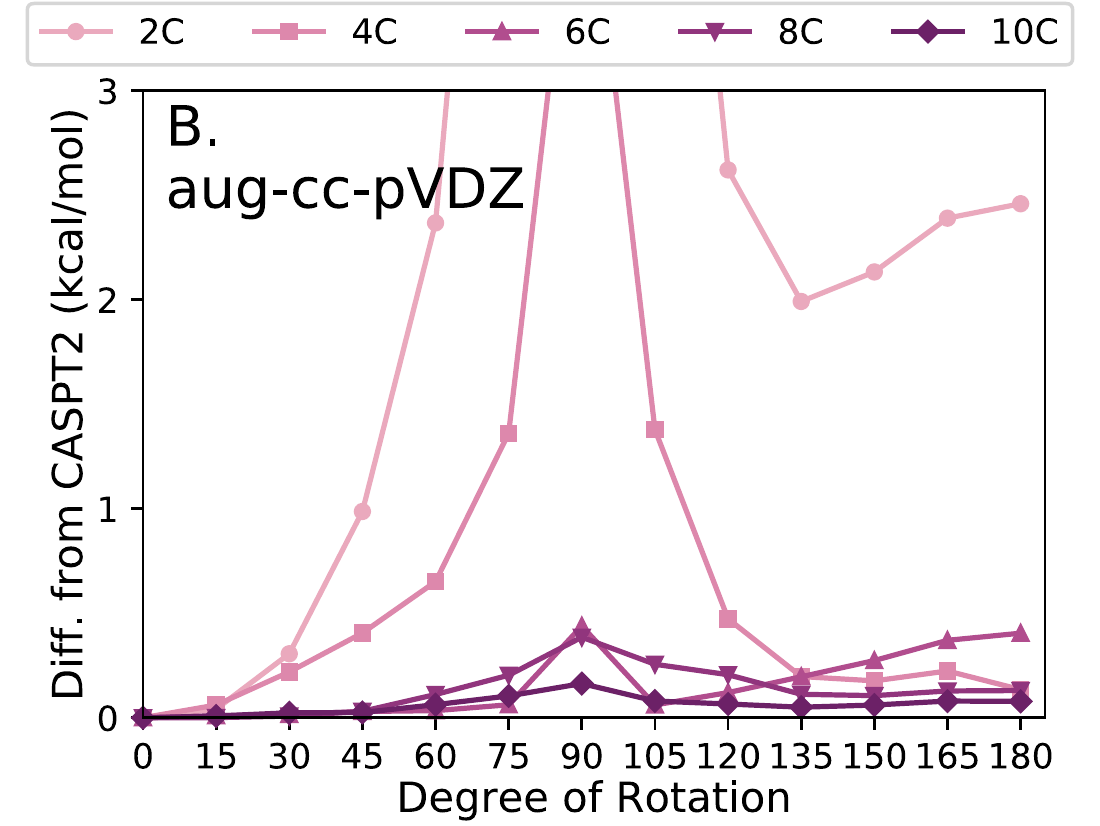}
    
    \vspace{8px}
    \includegraphics[width=0.5\textwidth]{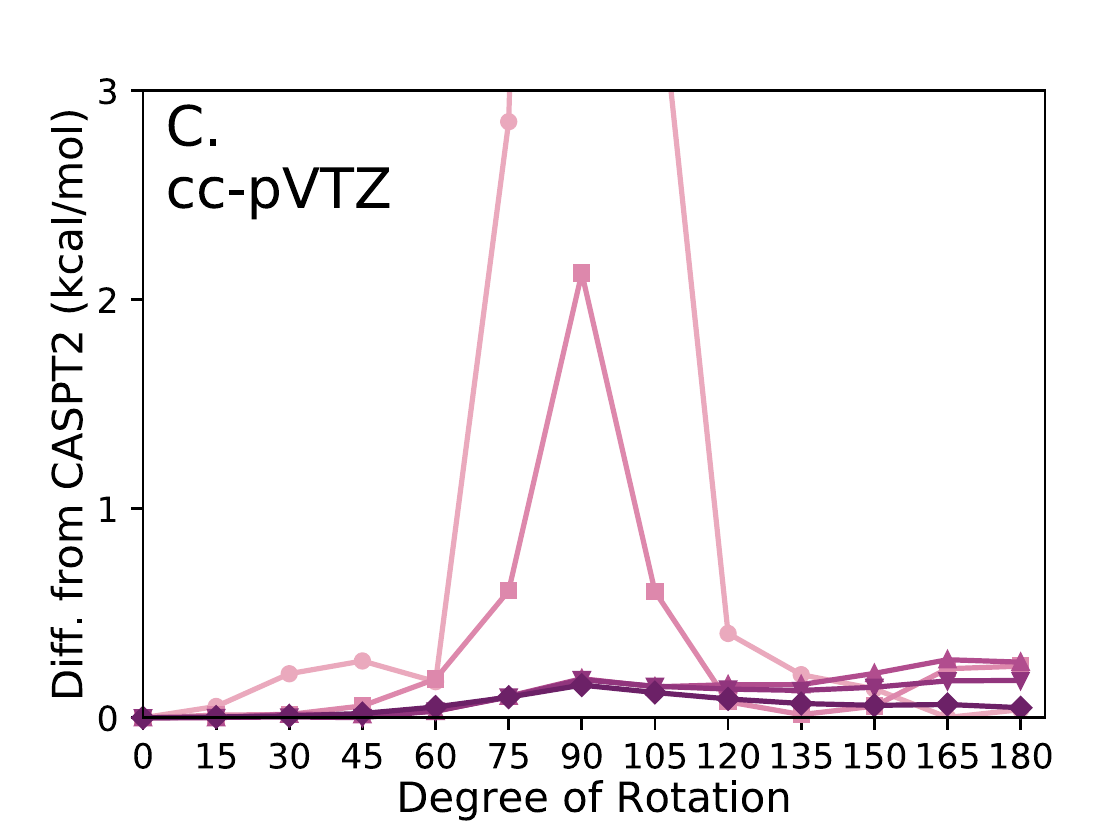}
    \caption{Absolute energy difference of rotation about a double bond calculated using CASPT2 embedded in M06\cite{Zhao2008} DFT exchange-correlation functional. At 90\degree\ rotation, pi bond is entirely broken and forms diradical. Sub-figures show: (A) WF subsystem division; (B) results of embedding with aug-cc-pVDZ; and (C) cc-pVTZ basis.}
    \label{fig:bond_rotation}
\end{figure}
Another potentially challenging multireference problem is the rotation of a system around a double bond, breaking the pi bond in the process to form a diradical (Figure \ref{fig:bond_rotation}A).
We applied our Huzinaga CASPT2-in-M06 embedding procedure to this system with the charging scheme identified in the previous section. 
Our results (Figure \ref{fig:bond_rotation}) illustrate convergence with the size of the WF system, in addition to convergence to the full WF result with only 6 carbon in the WF region.
Those systems with less than 6 carbon in the WF region only significantly deviate from full system results only at those regions of high multireference character, similar to the bond cleavage results.
These results demonstrate the robustness of the method with respect to multireference WF calculations.

\subsubsection{Fe-MOF-74 H\textsubscript{2} Adsorption}
One of the most important contributions of this method is the ability to calculate WF energies of incredibly large, complex systems. 
Given that our method can calculate multireference WF level energies and -- with the correct subsystem charging -- divide a system across complex interactions, we calculated the adsorption energy of hydrogen on a model cluster of Fe-MOF-74 for the singlet spin state. This small model has been used previously to represent the reaction center while remaining small enough to calculate full WF energies.\cite{Lee2014Design4} 
Initial CCSD(T) calculations on the model system indicated that a multireference WF method was necessary to adequately describe the adsorption interaction.

\begin{figure}
    \centering
    \includegraphics[width=.5\textwidth]{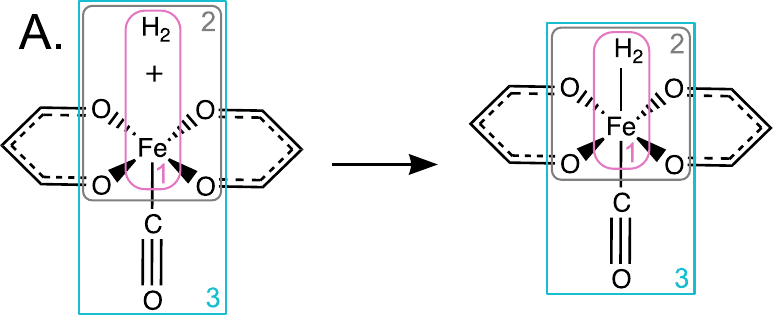}
    \includegraphics[width=0.5\textwidth]{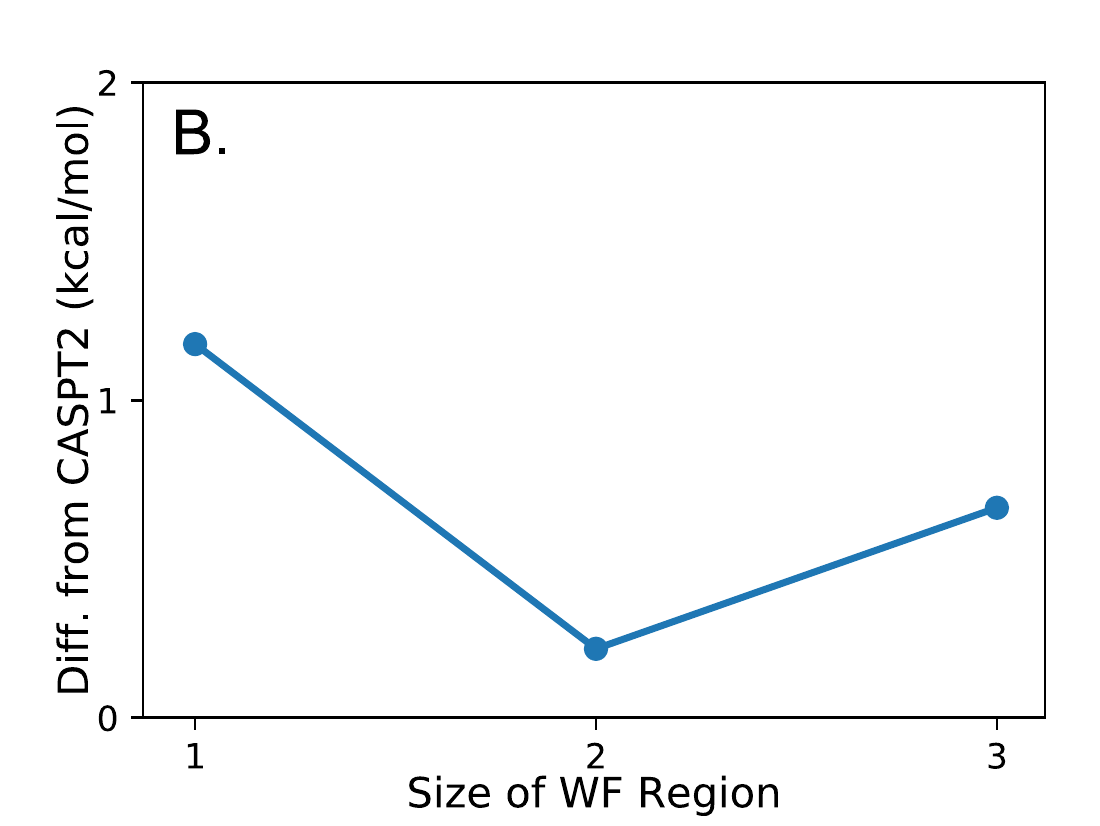}
    \caption{Subfigure (A) shows hydrogen gas adsorption on Fe-MOF-74. WF subsystems are specified with the smallest containing just the Fe and hydrogen adsorbant. Subfigure (B) shows absolute energy difference of hydrogen adsorption calculated using CASPT2 embedded in M06\cite{Zhao2008} DFT exchange-correlation functional and cc-pVDZ basis set.}
    \label{fig:femof74}
\end{figure}

As with the previous multireference calculations, we embedded CASPT2 within DFT using the M06\cite{Zhao2008} functional. 
Since the hydrogen adsorption is localized to the Fe metal center, we subdivided embedding systems as illustrated in Figure \ref{fig:femof74} and limited the CAS active space to 6 electrons in 5 orbitals for the bare MOF model and 8 electrons in 7 orbitals for the model with hydrogen bound. 
The active orbitals consist of the Fe 3d and bound hydrogen 1s orbitals. 
Figure \ref{fig:femof74}A shows similar convergence seen with other systems studied in addition to very close agreement to full system CASPT2 results with only the metal center and hydrogen within the WF subsystem (subsystem 1).  
This embedding scheme also serves to reinforce our previously established rule of thumb for charging systems. For this MOF system, subsystem 1 has a +2 charge in the WF region, corresponding with the oxidation state of the Fe, and a -2 charge of the DFT region to maintain neutrality.
Because there are no covalent bonds partitioned into subsystem 1, only ionic interactions, there is no need to move the electrons from a shared bond space.
Subsystems 2 and 3 have a -6 charge of the WF region and a +6 charge of the DFT region since subsystem division occurs across four conjugated carbon-oxygen covalent bonds.
Given our success when applying the method to this system, our embedding method is applicable to much larger, beyond current WF level calculations. 
We are actively applying our method to such systems.

\section{Conclusions}
The absolutely localized Huzinaga level-shift projection operator method of DFT embedding is an efficient, robust, and systematically improvable embedding technique. 
Across a diverse set of test systems, the Huzinaga embedding scheme consistently approached the full system WF calculation energy; replicating the energy within 1 kcal/mol for most systems with a fraction of the full system basis functions. Additionally across all systems, as the size of the WF region increased, the embedded energy approached the full system WF energy, demonstrating systematic improvability.
We also argue for the importance of balancing the size and flexibility of the WF subsystem with localized electron density leading to beneficial error cancellation.
By performing the WF calculation using only the subsystem basis functions, far fewer computational resources were needed to calculate energies of complex systems. As a result, accurate single reference and multireference WF energies may be calculated on systems that previously were too large for all but DFT methods, as demonstrated with the Fe-MOF-74 cluster model.  

Included in the Supplemental Information are all output files for the work presented here. This data set is also available at the Data Repository hosted at the University of Minnesota.\cite{Graham20192019GSEmbeddingPaperData} These output files also contain the input files used to generate the output.  Our QSoME code\cite{qsome2019} is open-source and requires an interface to the open-source PySCF\cite{Sun2018PySCF:Framework} program with an optional interface to Molpro\cite{Werner2012Molpro:Package, MOLPRO2019, Gyorffy2013AnalyticalFitting, Shiozaki2011Communication:Gradients, Celani2000MultireferenceFunctions}.  The input is very simple: definition of the subsytems, the charge associated with the subsystem, followed by the standard requirements for DFT and WF calculations (such as basis set, WF method, and exchange-correlation functional).  The combination of open-source code and simple input will allow researchers to apply this methodology to a wide range of applications and obtain high accuracy results at a significantly reduced computational cost.



\begin{acknowledgement}
This research was carried out within the Nanoporous Materials Genome Center, which is supported by the U.S. Department of Energy, Office of Basic Energy Sciences, Division of Chemical Sciences, Geosciences, and Biosciences under Award DE-FG02-17ER16362. The authors acknowledge the Minnesota Supercomputing Institute (MSI) at the University of Minnesota and the National Energy Research Scientific Computing Center (NERSC), a DOE Office of Science User Facility supported by the Office of Science of the U.S. Department of Energy under Contract No. DE-AC02-05CH11231, for providing resources that contributed to the results reported within this paper.
\end{acknowledgement}

\begin{suppinfo}
\end{suppinfo}

\bibliography{huzinaga2019Paper}
\begin{tocentry}
\includegraphics{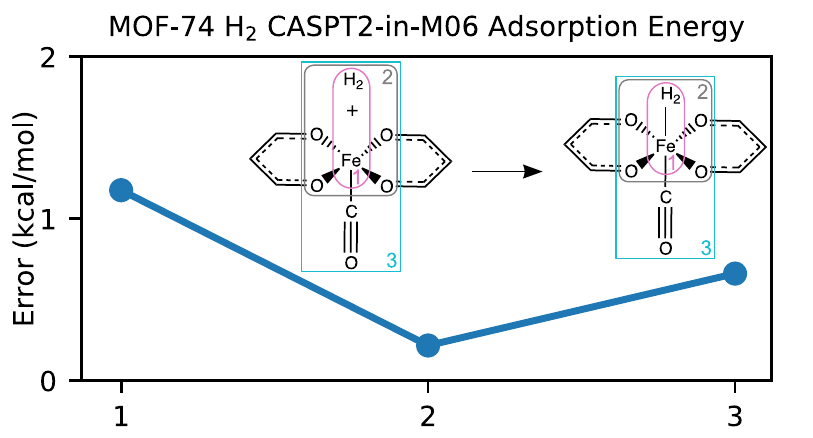}
\end{tocentry}
\end{document}